\documentclass[%
 reprint,
 amsmath,amssymb,
pra,
]{revtex4-2}

\usepackage{graphicx}
\usepackage{color}
\usepackage{dcolumn}
\usepackage{bm}
\usepackage{hyperref}

\begin{document}

\preprint{APS/PRA}

\title{Application of path-integral quantization to  indistinguishable particle systems topologically confined by a magnetic field}
\thanks{Published in Phys. Rev. A 97, 012108, 2018, \url{https://doi.org/10.1103/PhysRevA.97.012108}}

\author{Janusz E. Jacak}
 \email{janusz.jacak@pwr.edu.pl}
\affiliation{Deparment of Quantum Technologies, Wroc{\l}aw University of Science and Technology, Wybrze\.ze Wyspia\'nskiego 27, 50-370 Wroc{\l}aw, Poland}%



\begin{abstract}
We demonstrate an original   development of path-integral quantization in the case of a multiply connected configuration space of indistinguishable charged particles on a 2D manifold and exposed to a strong perpendicular  magnetic field. The system occurs to be exceptionally homotopy-rich  and the structure of the homotopy essentially depends on the magnetic field strength resulting in multi loop trajectories at specific conditions. We have proved, by a generalization of the Bohr-Sommerfeld quantization rule, that the size of a magnetic field flux quantum grows for multi loop orbits like $(2k+1)\frac{h}{c}$ with the number of loops $k$. Utilizing  this property  for  electrons on the 2D substrate jellium we have derived upon the path integration a complete FQHE hierarchy in excellent consistence with experiments. The path-integral has been next  developed to a sum over configurations, displaying various patterns  of trajectory homotopies (topological configurations),  which in the nonstationary case of quantum kinetics  reproduces some unclear formerly details in the longitudinal resistivity observed in  experiments. 
\end{abstract} 

\maketitle
\section{Introduction}
Path-integral formalism originally formulated for classical  real-valued stochastic processes \cite{wiener} in the form of an integral over histories, has been rediscovered  in a complex-valued  version  by Richard Feynman \cite{feynman1} and applied   to the definition  of the quantum propagator, being the matrix element of an evolution operator in the position representation. This breakthrough approach to quantization (called as a 'third formulation of quantum mechanics' besides the Schr{\"o}dinger
and Heisenberg ones) \cite{feynman1964,chaichian1,chaichian2} appears to be universal and crucial for development of almost all fields in modern quantum physics, field theory, unification of interactions and particle physics, condensed matter and statistical physics, superconductivity and phase transition theory, magnetism, gauge theories, quantum gravity, and even cosmology \cite{chaichian1,chaichian2,papad}. Despite the enormous success of the path-integral quantization, its precise mathematical formulation is still developing because the measure in a path space does not fulfill the conditions for Lebesgue measure  \cite{kac,papad}. Some insufficiency in mathematical rigor has been, however, complemented with continuous  advances in mathematical formulation, cf., e.g., Ref. \onlinecite{dewitt-m}.

An important progress in application of the Feynman path quantization (FPQ) has been achieved also in the case of topologically non trivial spaces including multiply connected  spaces. The latter are identified by the fundamental group ($\pi_1$ homotopy group) \cite{spanier1966,mermin1979,rider}, which for a simply connected space is the trivial group but is not a trivial one  for any multiply connected space \cite{lwitt}. It has been proved that in the case a multiply connected space  the path-integral must be developed by adding a summation over the elements of the fundamental group of this space with particular component contributions assigned by one-dimensional unitary (due to causality) representation (1DUR)  of the fundamental group elements \cite{szulm,lwitt} and simultaneously taken the trajectory in the action in the exponent in the path-integral with additional loops from $\pi_1$. This is caused by the fact that for a multiply connected space its fundamental group, $\pi_1$, displays non homotopic (not continuously transformable) path loops and the domain of the path-integral decomposes into disjoint segments enumerated by $\pi_1$ elements. The contributions of all these segments must  be included to the total path-integral by summation because due to linearity the total path-integral decays into a sum over non homotopic path space segments with an unitary weight for each component (just defined by a 1DUR of $\pi_1$).

Originally, in Ref. \onlinecite{szulm} it has been demonstrated that the $\pi_1$ group multiply connected space  SO(3)  treated as some internal space  of a single-particle in 3D may support spin upon Feynman path-integrals, which was next generalized, in Ref. \onlinecite{lwitt}, to any multiply connected space related to particle trajectories. Scalar unitary  representations (1DURs) of  $\pi_1$  group of such space may assign  unitary weights for different non homotopic contributions to the path-integral. To the class of multiply connected spaces  belong also the multi-particle configuration spaces  of identical quantumly indistinguishable particles. The fundamental groups for $N$-particle configuration spaces are called full braid groups \cite{wilczek,wu,birman}. They display a topology constraints imposed by  various manifolds on which particles are located when considering the homotopy equivalence or non-equivalence of trajectories traversed by particles  interchanging positions on these manifolds \cite{birman,jac-ws}. Different 1DURs of braid groups assign thus different quantum statistics of identical indistinguishable quantum particles corresponding to their classical counterparts \cite{lwitt}.   It has been proved \cite{birman,wu,jac-ws} that for any many-particle system in 3D space (and on higher dimensional manifolds) the full braid group is always the permutation group, $S_N$ ($N$ is the number of particles). The permutation group $S_N$ has only two 1DURs defining two possible quantum  statistics in 3D, bosonic and fermionic ones. Such an development of the  path-integral quantization from its pristine form for a single-particle \cite{feynman1964}, for the case of  systems with many indistinguishable particles, in a natural way introduces the nontrivial fermionic statistics in purely topological terms and supplements the path-integral reformulation  in Grassmann variables for fermions \cite{bere}. More spectacular is an application of FPQ approach to systems of many indistinguishable particles confined to a planar manifold \cite{wu} or locally planar, like a sphere or a torus \cite{jac-ws,einarsson}. In these cases the braid groups are not permutation groups but are rather infinite discrete groups with  rich unitary representations  assigning so-called anyons   besides bosons and fermions and related fractional statistics \cite{wu,wilczek,imbo}.  An especially interesting opportunity for further development of FPQ arises, however, in 2D systems of electrons  in the presence of quantizing strong perpendicular magnetic field which additionally and considerably modifies the topology of planar systems. Such  systems  were intensively investigated both experimentally and theoretically since the early 80s of the past century in context of an integer quantum Hall effect (IQHE)  and a fractional quantum Hall effect (FQHE). The interest in IQHE and FQHE has been renewed more recently due to experimental advances  in graphene. A current Hall experiments with graphene  continuously supply new  portions of information on unconventional 2D  physics in this material. An application of the  path-integral approach to the field of 2D Hall physics significantly enhances  the transparency of a related theory and contributes to its development \cite{wu,jac-ws}. Moreover, an analog of FQHE in a topological Chern insulator \cite{hasan,ftci1,ftci2} when the magnetic field has been substituted by a Berry field has been demonstrated \cite{hasan}. 

In this paper, we present a new important homotopy aspect of the path-integral quantization of a 2D charged system upon the magnetic field presence allowing for an  elucidation of the topological nature of FQHE and related so-called composite fermions \cite{epl,atmp,jac-ws}, introduced formerly in a heuristic manner in order to describe the unconventional  quantum physics in 2D \cite{jain}. We have developed here  the FPQ approach toward characterization of  electron correlations in FQHE being out of reach for a conventional theory, but which can be understood in homotopy terms and  are   experimentally  observed  in FQHE hierarchy for GaAs 2DEG and for graphene \cite{pan2003,nat,nature}. Moreover, we propose a new method called by us as the summation over braid configurations in the path-integral in a non-stationary case, which  beneficially meets with some characteristics in the  transport effects \cite{pan2003} observed  in Hall systems but was overlooked in the conventional local-type theory. 

The paper is organized as follows: in the next paragraph, we present the general idea of the modification of braid groups in the case of 2D multiparticle charged systems upon a strong perpendicular magnetic field called the cyclotron braid subgroup approach. In the following paragraph we apply this concept to an explanation of fractional quantum Hall phenomena out of reach for conventional composite fermion model but observable in the experiments. This paragraph is followed by the next one in which we present a new concept of summation over topological configurations essential for a description in terms of path-integrals of some kinetical nonstationary  characteristics in FQHE like the longitudinal conductivity, which was not formerly explained. Next, we present a generalization of the Bohr-Sommerfeld rule in a homotopy-rich case, which proves a size growth of multi loop 2D orbits in comparison to single-loop ones. The comparison with experiment is included  exhibiting an increase in consistence of the theory with  experimental data  and the theory transparency.

\section{Homotopy corrections to  path-integral for identical quantumly indistinguishable particles}

\begin{figure*}[ht]
\centering
\makebox[\linewidth]{
\scalebox{1.0}{\includegraphics{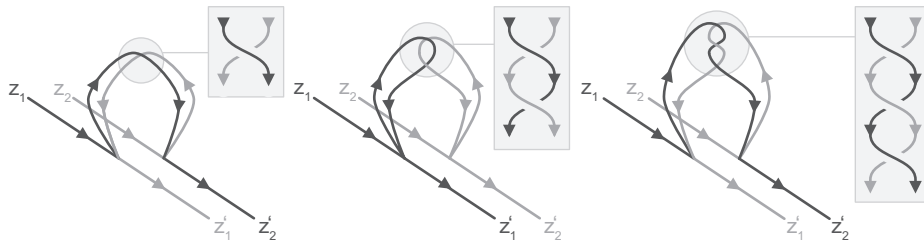}}
}
\caption{\label{fig25} Example of non homotopic trajectories obtained by addition of various non homotopic braids to two-particle trajectory.}
\end{figure*}

  The method utilizes the Feynman path-integral quantization formalism appropriately lifted to describe multiparticle systems assuming particle indistinguishability\cite{wu,wilczek,lwitt}.   The latter property is incorporated into  the definition of a configuration space for $N$ particles on the manifold $M$ in the following form, $ \Phi_N(M)=(M^{N}-\Delta)/S_N$, where $M^N$ is an $N$-fold normal product of $M$, $\Delta$ is the subset of diagonal points in $M^N$ (if at least two particle positions coincide), removed to assure particle number conservation. The quotient structure by the permutation group $S_N$ accounts for particle indistinguishability. The  loops  of multiparticle trajectories joining initial and final particle positions distinct only in their   enumeration (unified, however, due to their indistinguishability) have the form of closed braids, i.e., of closed bunches of  braided single-particle trajectories impossible to disentangle. Such loops (braids)  can be adjoint to any point of an open multiparticle trajectory in the path-integral for a many-particle system  in its pristine version, without any topological effects included \cite{feynman1964,chaichian1,chaichian2}. The open path   connects some  point in $\Phi_N(M)$ denoted by particle coordinates, $z_1,\dots, z_n$ (the initial point at time instant  $t$) with another point $z'_1,\dots,z'_N$ (the final point at time instant $t'$) in $\Phi_N(M)$.  The braid  loops  in $\Phi_N(M)$  are, however, not equivalent topologically and form disjoint homotopy classes. These classes create the so-called full braid group $B_N(M)$ being the first homotopy group $\pi_1(\Phi_N(M))$.    Because braid loops are mutually non homotopic, the resulted open trajectories with adjoint distinct braids are also topologically  inequivalent (as the braids and the whole trajectories with adjoint braids as well cannot be transformed one into another in a continuous way---as illustrated on the simple example in  Fig. \ref{fig25}). All  trajectories with loops fall thus  to disjoint classes of non homotopic trajectories (which cannot be unified or mixed  by continuous deformations). These classes form together  the complete domain of the Feynman path-integral,

\begin{equation}
\begin{split}
I(z_1&,\dots, z_N, t; z'_1,\dots, z'_N, t')\\
&=\sum_{l\in \pi_1(\Omega)}e^{i\alpha_l}\int d\lambda_l e^{iS[\lambda_l(z_1,\dots, z_N, t; z'_1,\dots, z'_N, t')]/\hbar},\\
\end{split}
\label{path}
\end{equation}

\noindent  where, $I(z_1,\dots, z_N, t; z'_1,\dots, z'_N, t')$ is the propagator, i.e., the matrix element of the evolution operator of the total $N$-particle system in the position representation which determines the probability amplitude (complex one) of quantum transition from the point, $z_1,\dots, z_N$, in time instant $t$ to the other point in the configuration space, $z'_1,\dots, z'_N$,  in time  instant $t'$, $d\lambda_l$ is the measure in the path space sector enumerated by the braid group $\pi_1(\Phi_N(M))$ elements  (any braid group is  always countable).  $S[\lambda_l(z_1,\dots, z_N, t; z'_1,\dots, z'_N, t')]$ is the classical action for the trajectory $\lambda_l$ joining selected  points in the configuration space $\Phi_N(M)$ between time instances $t$, $t'$ and lying in $l$th sector of the trajectory space with  $l$th braid loop adjoint. The whole space of trajectories is decomposed into disjoint sectors enumerated by the braid group element discrete index $l$ (braid groups are countable).  The discontinuous decomposition of the domain of the path-integral into disjoint sectors (topologically inequivalent) precludes a definition of the path measure $d\lambda$ uniformly on the whole space of paths (due to the continuity constraint) and for each sector the measure $d\lambda_l$ must be defined separately and finally the  contributions of all sectors must be summed with  unitary weights $e^{i\alpha_l}$ (unitarity is caused by the causality constraint). It has been proved \cite{lwitt} that these unitary factors establish a one dimensional unitary representation  (1DUR) of the full  braid group.   Distinct unitary  weights   in the path-integral  (i.e., distinct 1DURs of the braid group) determine different sorts of quantum particles corresponding to the same classical ones. Braids describe particle  exchanges, thus their  1DURs assign quantum statistics. Equivalently, the 1DUR  of  a particular braid defines a  phase shift of the multiparticle wave function $\Psi(z_1,\dots ,z_N)$ when its arguments $z_1, \dots, z_N$ (classical coordinates of particles on the manifold $M$) mutually exchange themselves according to this  braid \cite{sud,imbo} (let us emphasize that  these exchanges are {\it not}  permutations and the path is important, unless the manifold $M$ is three or higher dimensional space without linear topological defects, like strings \cite{birman,sud,imbo}).

 All  quantum multiparticle correlated states (including correlated states related to IQHE or FQHE)  must be thus characterized  unavoidably   by a certain 1DUR of the full  braid group for a particular system. In 3D, the full  braid groups of  $N$ particle systems are  always the $N$-element permutation groups (regardless of charge, interaction, or magnetic field presence). There exist only two 1DURs for an arbitrary permutation group: $\sigma_j\rightarrow \left\{
\begin{array}{l}
e^{i0}=1\\
e^{i\pi}=-1\\
\end{array}\right.$,
where $\sigma_j$, $j=1,\dots, N-1$, the generators of the braid group are braids (multiparticle classical trajectory bunches) for exchanges of positions of $j$th particle with $(j+1)$th particle, when other particles remain at rest. 1DUR=1 defines bosons and 1DUR=$-1$ defines fermions in 3D multiparticle systems. 
For 2D multiparticle systems the braid groups are essentially different than the permutation groups \cite{jac-ws,wu,birman} and their  1DURs are different as well, $\sigma_j\rightarrow e^{i\alpha},$ $\alpha\in(-\pi,\pi]$. Various 1DURs define 2D anyons (including 2D fermions for $\alpha=\pi$ and bosons for $\alpha=0$) at the absence of a quantizing magnetic field. 

\subsection{Braid groups for 2D electrons in the presence of a strong magnetic field}

In a 2D manifold (let us assume here $M=R^2$ plane) for charged repulsing electrons in the presence of a strong perpendicular magnetic field the braid group approach does not resolve itself to anyons only! A strong magnetic field perpendicular to the basal plane itself changes  significantly the path homotopy and modifies the full braid group structure, which appears to be that topological factor conditioning FQHE manifestation in the planar Hall configuration  according to the same general  scheme  even in completely different systems with different single-particle properties as is actually observed in experiment for conventional GaAs 2D systems \cite{prange}
 as well as in a quasi relativistic graphene monolayer \cite{dean2011,amet,feldman,feldman2,nat} or bilayer \cite{amet,bil1,nature}.

The common property for all 2D charged systems at a strong enough magnetic field is that the planar  quantized   orbits may be shorter than the particle separation. This is important   when particles are   uniformly distributed on the plane with classical positions fixed by the Coulomb  repulsion of electrons. The classical distribution of 2D charged repulsing particles on the uniform jellium  at $T=0$ K is the static triangular Wigner lattice. If the too short  cyclotron orbit does not match neighboring particles then this  precludes the existence of the braid group generators $\sigma_j$, i.e., exchanges of neighboring particles are impossible.  Let us recall that braids are multiparticle trajectory loops in {\it classical} configuration space $\Phi_N(M)$ where the topology of $M$ decides the homotopy properties of these  trajectories. The  braids $\sigma_j$,  which for charged 2D particles in the presence of a magnetic field {\it  must} be built from pieces of classical cyclotron orbits,  cannot  be defined in the case when cyclotron orbits do not fit to particle separation rigidly fixed in the classical Wigner crystal. Too short $\sigma_j$ braids---elementary exchanges of neighbors---cannot be implemented as cyclotron orbits do not reach neighboring particles. Thus $\sigma_j$ must be  rejected from the braid group. Nevertheless, it has been  proved \cite{jac-ws,epl} that remaining in the braid group other  braids are on the 2D manifold large enough to match neighboring particles and these remaining, sufficiently large  braids form the subgroup of the original group. This subgroup is  called  the cyclotron braid subgroup \cite{jac-ws}.  The generators of the cyclotron subgroups are multi loop braids---such braids have in 2D larger size than single-loop braids \cite{epl} (the formal proof of this fact is placed in Sec. \ref{10} of the present paper). Hence, the cyclotron subgroups allow for the definition of quantum statistics in the presence of a strong magnetic field via their 1DURs in the path-integral formalism upon the same scheme presented in the previous paragraph, however, with the distinction that the summation over braid group elements does not concern now the full braid but its cyclotron braid subgroup. This is an important difference which admits identification of new quantum particles---composite fermions (and more generally, composite anyons including composite bosons), despite the 1DURs for the cyclotron subgroup defined on its generators may coincide with the original full braid group representations. The 1DURs for the cyclotron braid subgroups are as follows:
 
\begin{equation}
\label{phase}
b_j=\sigma^q_j\rightarrow e^{i q\alpha},\;\alpha\in(-\pi,\pi],\;q\text{-odd},
\end{equation}

\noindent where $b_j=\sigma_j^q$ are generators of the $q$-type cyclotron subgroup defining exchanges of neighboring $j$th and $(j+1)$th particles with $\frac{q-1}{2}$ additional loops \cite{epl,jac-ws}. The 1DUR (\ref{phase}) actually may sometimes coincide with the original full braid group representation, $\sigma_j\rightarrow e^{i\alpha}$, as it is noticeable e.g., for $\alpha=0$  or $\pi$, but the difference in the domain of the path-integral still remains. The quantum difference between ordinary fermions ($\alpha=\pi$) and composite fermions  resolves itself to the distinct domain for the homotopy class summation in the Feynman path-integral (\ref{path}). In the case of $\alpha \neq 0,\pi$ 1DURs (\ref{phase}) related to composite anyons is different than $e^{i\alpha}$ characterized anyons. Moreover, even in the case when $\alpha =\pi$, the phase shift $q \pi$ in (\ref{phase}) is different than $\pi$ and assigns the statistics originally identified by Laughlin in his famous function  \cite{laughlin2}.

The essential for cyclotron braid structure fact that in 2D multi loop orbits have a larger size in comparison to single-loop orbits can be proved independently by application of the Bohr-Sommerfeld quantization rule in a homotopy-rich case as is presented in Sec. \ref{10}. Worth noting is that this quasiclassical proof holds for any interaction between particles.       

1DURs of various  cyclotron braid subgroups generated by  fermionic 1DUR of initial full braid group, i.e., $\alpha=\pi$,  define specific types of composite fermions and allow for construction of related multiparticle trial wave functions for FQHE using symmetry constraints precisely defined by the form of cyclotron braid subgroup generators as will be demonstrated in the subsequent paragraph according to rules relating braid  features with wave the function symmetries \cite{sud,imbo}. Determined in this way,  wave functions  (here without the need of an artificial  projection onto the LLL, whih was inherent to the phenomenological Jain's idea of conventional composite fermions  \cite{jain2007}) pretty well  agree with  the exact diagonalization on small models and with the experiment---as will be illustrated in the next paragraph. Composite fermions are thus not equipped with auxiliary flux quanta but acquire the needed Laughlin-type phase shift \cite{laughlin2} according to the  1DUR of the cyclotron subgroup  generated by multi loop braids. This proves that the composite fermions [for $\alpha=\pi$  in Eq. (\ref{phase})] and, more general, the composite anyons (for arbitrary $\alpha \in (-\pi, \pi]$) are not any quasiparticles dressed with interaction (like Landau quasiparticles in solids \cite{abrikosov1975}) but are rather different types of quantum particles conditioned by the topological homotopy constraints imposed on charged interacting particles in 2D upon a sufficiently strong magnetic field.  The trial wave functions fp not need to be  projected from  higher LLs (as in the conventional heuristic composite fermion model \cite{jain2007})  but are uniquely defined according to symmetry imposed on the holomorphic function  by the appropriate 1DUR  of the particular  cyclotron braid group generators  adjusted to the commensurability between the cyclotron orbit size and the particle separation. 

How to adjust this commensurability? The answer resolves  itself to the specific to 2D manifold property that the external magnetic flux passing a planar multi loop orbit must be divided between all loops. Eventually, per each loop falls only fraction of the flux like for a smaller field and its size effectively grows allowing to fit the interparticle separation exceeding the single-loop cyclotron orbit size, as has been also formally proved by the quasiclassical Bohr-Sommerfeld quantization rule (cf. Sec. \ref{10}).  Each loop in the multi loop orbit can be accommodated to particle separation including nearest and next-nearest neighbors individually upon various patterns adjusted to the particle density fixed by the filling fraction of the Landau level (LL). The filling fractions $\nu=\frac{N}{N_0}$ ($N$ is the constant particle number, $N_0=\frac{BSe}{h}$ is the LL degeneracy, $S$ is the sample surface) vary with magnetic field and some of them are featured  by the cyclotron commensurability condition. This criterion discriminates the majority of filling fractions (including all irrational ones) but selects the specific hierarchy of FQHE at some magic-looking rational fractional fillings as visible in the experiment.       

From this point of view, it is unimportant and irrelevant to look for FQHE filling hierarchy from single-particle properties of a particular  Hall system, like a pseudo-relativistic LL structure of monolayer or bilayer graphene, otherwise completely different than in a conventional GaAs. single-particle band properties are unimportant for homotopy features essential for FQHE and its  hierarchy of fillings unless they can change the topology (such a change happens, however, in a bilayer graphene when the interlayer tunneling of electrons dismisses strict 2D topology and substitutes it with a specific different bilayer topology \cite{nature}). The topology is immune to the dynamics particularities including also  particle electric interaction. The Coulomb interaction defines, however, the initial uniform classical Wigner crystal  distribution of  particles---the start point and arena for braid definition being the essential  prerequisite for commensurability constraints imposed by cyclotron braids---therefore the Coulomb interaction plays a central role in FQHE formation. The FQHE hierarchy is repeated in various systems in a similar form despite single-particle band structure differences and even in quite different systems like in the topological Chern insulator \cite{hasan,ftci1,ftci2}, which supports a  topological conditioning of this hierarchy.  

The construction of appropriate cyclotron braids is possible  only at some specific filling rates of the LL when the commensurability constraints imposed on the size of particular loops of multi loop cyclotron orbit  versus particle separation including next-nearest neighbors can be  fulfilled. Each loop of the cyclotron multi loop orbits which  built braids must fit to the interparticle separations.  The  discrimination of the filling rates by this commensurability condition  results in filling  hierarchy in fully consistence with the  experimental observations of FQHE hierarchy in contrast to conventional model of composite fermions \cite{jain2007} which failed in more complicated homotopy situations. The model od composite fermions with auxiliary field flux quanta attached to electrons \cite{jain} may be treated as an effective model for multi loop orbits, but only in the simplest case of the homotopy and orbit commensurability. The homotopy braid group approach upon the scheme of Feynman path-integral is mathematically rigorous and  more general, it reproduces also fractions experimentally observed in conventional semiconductor 2DEG which are out of the conventional composite fermion series (both in the lowest LL (LLL) and in higher LLs \cite{nature}.  The conventional composite fermion  model agrees with the braid group approach for the simplest case of the commensurability only (in case precisely defined in paragraph \ref{2}). The short summarizing  of the topological cyclotron braid subgroup approach to FQHE  in the LLL is given below.

\subsection{Cyclotron braid commensurability  for FQHE states in the LLL of GaAs Hall system}
\label{2}
\begin{figure*}[ht]
\centering
\resizebox{1\textwidth}{!}{\includegraphics{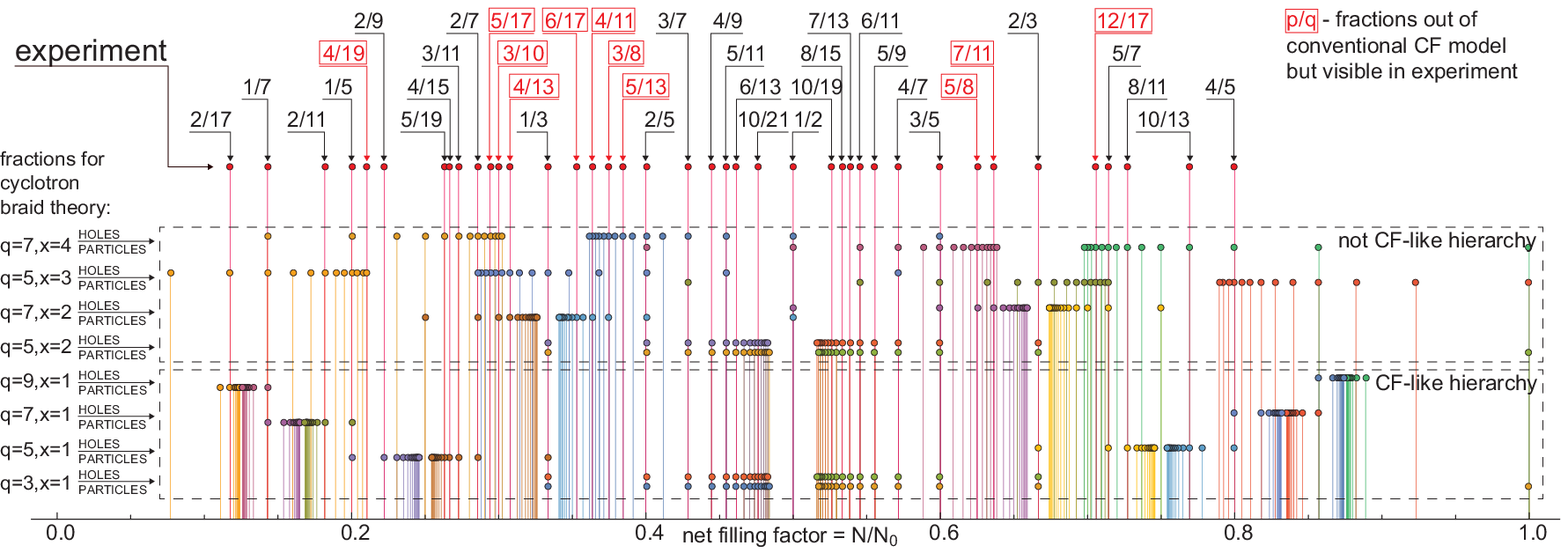}}
\caption{\label{fig1} Comparison of the hierarchy (\ref{444}) with all measured fractional filling rates for FQHE features in the LLL (spin polarized). The hierarchy series acc. (\ref{444}) for several $y$ each are displayed, filling rates beyond the conventional  hierarchy of  Jain's composite fermions are shown in red frames (Hall metal state fraction 1/2 is marked).}
\end{figure*}

 One can identify  the correlated states at fractional fillings generalizing the genuine pattern of the correlation of IQHE, $\frac{S}{N}=\frac{S}{N_0}$ when cyclotron orbit size $\frac{S}{N_0}$ ($N_0=\frac{eBS}{h}$ is the LL degeneracy) fits to electron separation $\frac{S}{N}$.  At fractional fillings of the LLL the cyclotron orbits $\frac{h}{eB}$ are smaller than $\frac{S}{N}$  and cyclotron orbits cannot match neighboring electrons. For establishing of any correlated state, the particle exchanges are, however, necessary to define statistics of quantum particles via a choice of a braid group 1DUR in the path-integral. Exclusively in 2D, the multi loop cyclotron orbits have larger size in comparison to single-loop ones at the same magnetic fields \cite{jac-ws,epl} (cf. the  proof in paragraph \ref{10}). It follows from the  distribution of the external field $B$ flux attributable per particle among all loops of the multi loop cyclotron orbit all located, however,  in the same plane. The condition for commensurability   attains thus the  general form including matching by multi loop orbits  of nearest or next nearest electrons:

\begin{equation}
\label{222}
\frac{BS}{N}=(q-1)\frac{h}{ex}\pm\frac{h}{ey},
\end{equation}

\noindent where:
$q$ is the number of loops of single cyclotron orbits ($q$ must be odd integer in order to ensure the corresponding braid to describe particle exchange---the braid generator with $n$ additional loops corresponds to $2n+1=q$-loop cyclotron orbit \cite{epl,jac-ws}). At magnetic fields in 2D the braids are built from half-pieces of cyclotron orbits provided that these orbits accurately fit to neighboring  (nearest or next-nearest) particle separation  at the uniform particle distribution caused by the electric repulsion. In condition (\ref{222})
$x\geq 1$ (integer) indicates the commensurability of $q-1$ single loops from the $q$-loop cyclotron orbit to every $x$th particle on the plane ($x=1$ corresponds to nearest neighbors, whereas $x>1$ to next-nearest ones);
$y\geq x$ (also integer) indicates the commensurability of the last loop of the $q$-loop orbit with every $y$th particle (next-nearest neighbors if $y>1$);
$\pm$ indicates the same or opposite (of eight-figure-shape) orientation of the last i.e., $q$th loop.
From (\ref{222}) we obtain the following conditions, 

\begin{equation}
\label{444}
\begin{array}{l}
\nu=\frac{N}{N_0}=\frac{xy}{(q-1)y\pm x},\text{ for band electrons},\\
\nu=1-\frac{xy}{(q-1)y\pm x},\text{ for band holes},\\
\end{array} 
\end{equation}

\noindent for the general hierarchy of correlated states in the LLL describing the FQHE hierarchy. For $x=1$ the hierarchy (\ref{444}) reproduces the conventional composite fermion hierarchy. For $x > 1$ the  hierarchy (\ref{444})  is beyond the ability of the Jain's model of composite fermions \cite{jain} and displays  filling ratios for FQHE in the LLL including those  outside the Jain's hierarchy, which are, however,  observed in the experiment in GaAs 2DEG \cite{pan2003}. The detailed comparison with the experimental data is summarized in Fig. \ref{fig1}. 

 The Jain's composite fermion  model agrees with the simplest commensurability case ($x=1$) and breaks down in more complicated commensurability instances as given by Eq. (\ref{444}) for $x>1$. In Fig. \ref{fig1} in  red frames, filling rates out of main Jain's composite fermion hierarchy are indicated, but visible in experiments and successfully reproduced by the hierarchy (\ref{444}).

\begin{figure*}[ht]
\centering
\resizebox{0.9\textwidth}{!}{\includegraphics{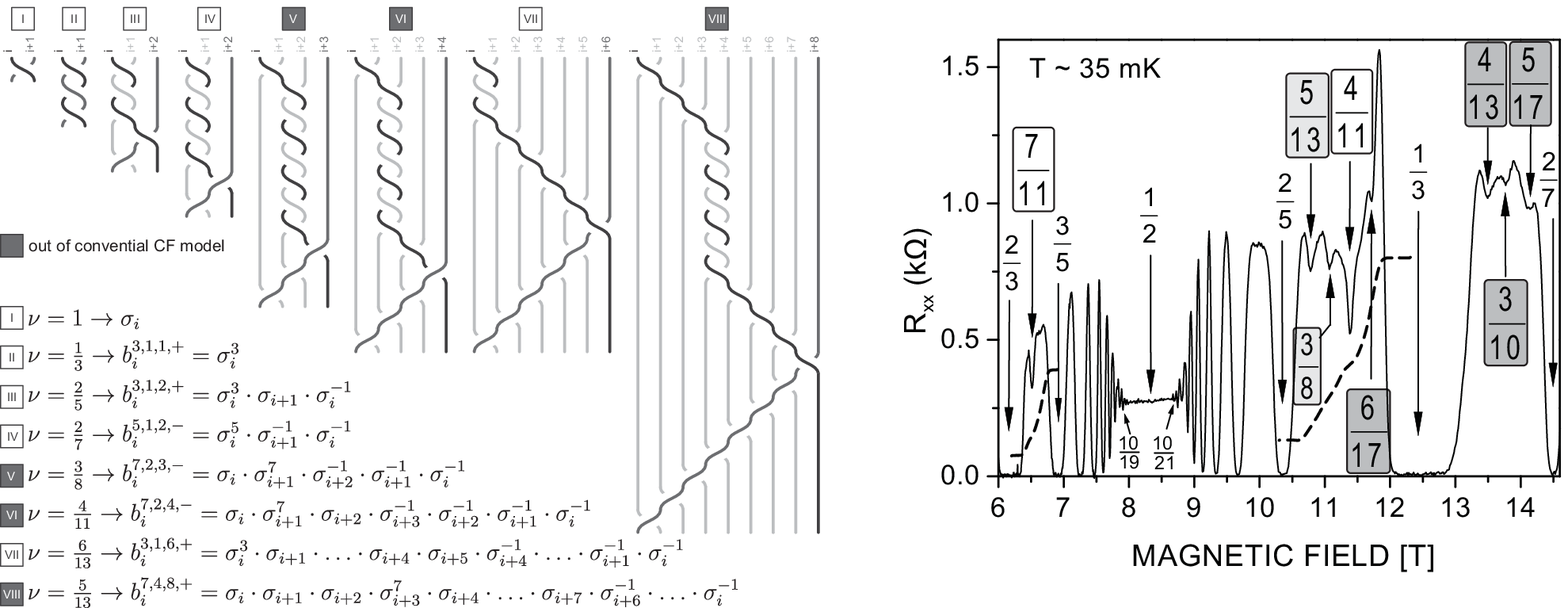}}
\caption{\label{fig3}In left panel,  the braid cyclotron subgroup generators for several exemplary filling fractions (examples of generators for filling fractions which cannot be derived using the conventional composite fermion  model are marked). In right panel, the longitudinal resistivity measured in GaAs 2DEG (after Ref. \onlinecite{pan2003}) with marked similar levels of residual resistivity corresponding to similarly not fully correlated electrons for fractions out of the conventional composite fermion model, in compliance with the related homotopy configurations.}
\end{figure*}

The limit $y\rightarrow \infty$ displays the hierarchy of the Hall metal exactly in the same manner as for the archetype of Hall metal at $\nu=1/2$ (the last orbit is then infinite and fits to infinitely distant particles as in the normal Fermi liquid without any magnetic field). The general Hall metal hierarchy in the LLL has thus the form:

\begin{equation}
\label{555}
\begin{array}{l}
\nu= \frac{x}{q-1},\text{ for electrons},\\
\nu=1-\frac{x}{q-1},\text{ for holes}. \\
\end{array}
\end{equation}

\noindent
Note that Hall metal correlation can manifest itself at fractions not necessarily with even denominators (for $x>1$ even, beyond the conventional Jain's composite fermion  concept), similarly as the hierarchy (\ref{444}) displays fractions both with odd and even denominators in compliance with the experimental observations \cite{pan2003}. Some fractions are repeated in various lines of the general hierarchy (\ref{444}). This fact reveals the possibility of various types of commensurability of multiloop cyclotron orbits with interparticle spacing $\frac{S}{N}$. The advantage of one commensurability over the others (alternative ones at the same filling ratio) is related with energy minimization, i.e., with the minimization of the Coulomb interaction. 

\subsection{Trial wave functions for FQHE states in the LLL for GaAs 2DEG}
\label{3}
For the simplest line of the hierarchy (\ref{444}) with $x=y=1$, i.e., $\nu=\frac{1}{q}, \;q\text{-odd}$, the corresponding wave function has been given by Laughlin in the form \cite{laughlin2}:

\begin{equation}
\label{666}
\Psi_q(z_1,z_2,\dots, z_N)=A\prod\limits_{i,j,i>j}^{N,N}(z_i-z_j)^q e^{-\sum\limits_i^N\frac{|z_i|^2 }{4l^2}},
\end{equation}

\noindent where $z_i=x_i+iy_i$ is $i$th particle classical position on the complex plane (the argument of the quantum multiparticle wave function), $l=\sqrt{\frac{h}{eB}}$ is the magnetic length, and the product $\prod\limits_{i,j,i>j}^{N,N}(z_i-z_j)^q$ is the Jastrow polynomial, $A$ is an appropriate normalization constant. The defining characteristic of the Laughlin function is that the $q$-fold zero at each  particle  keeps particles apart, and thus diminishes the Coulomb interaction energy.
The function (\ref{666}) must transform itself according to the 1DUR of the  cyclotron braid subgroup with generators $\sigma_i^q$. And indeed, for the 1DUR of the full braid group , $\sigma_i\rightarrow e^{i\alpha}$ with $\alpha =\pi$ (fermionic) one gets from Eq. (\ref{phase}) $e^{iq\pi}$ as the 1DUR of $\sigma_i^q$, which coincides with the  Laughlin  phase.

For the hierarchy (\ref{444}) the generators (describing elementary exchanges) of the appropriate more complicated cyclotron braid subgroups are defined as follows (for $\pm $ in (\ref{444})):

\begin{widetext}
\begin{equation}
\label{777}
\begin{array}{l}
b_{i}^{q,x,y, +}=(\! \sigma_i\! \cdot\! \sigma_{i+\!1}\! \cdot\! ...\! \cdot\! \sigma_{i+x-2}\! \cdot\! \sigma_{i+x-\!1}
\! \cdot\! \sigma_{i+x-2}^{-\!1}\! \cdot\! ...\! \cdot\! \sigma_{i+\!1}^{-\!1}\! \cdot\! \sigma_{i}^{-\!1})^{q-\!1} \cdot\sigma_i\! \cdot\! \sigma_{i+\!1}\! \cdot\! ...\! \cdot\! \sigma_{i+y-2}\! \cdot\! \sigma_{i+y-\!1}\! \cdot\! \sigma_{i+y-2}^{-\!1}\! \cdot\! ...\! \cdot\! \sigma_{i+\!1}^{-\!1}\! \cdot\! \sigma_{i}^{-\!1},\\
\text{and}\\
b_{i}^{q,x,y, -}=(\! \sigma_i\! \cdot\! \sigma_{i+\!1}\! \cdot\! ...\! \cdot\! \sigma_{i+x-2}\! \cdot\! \sigma_{i+x-\!1}\! \cdot\! \sigma_{i+x-2}^{-\!1}\! \cdot\! ...\! \cdot\! \sigma_{i+\!1}^{-\!1}\! \cdot\! \sigma_{i}^{-\!1})^{q-\!1} \cdot (\! \sigma_i\! \cdot\! \sigma_{i+\!1}\! \cdot\! ...\! \cdot\! \sigma_{i+y-2}\! \cdot\! \sigma_{i+y-\!1}
\! \cdot\! \sigma_{i+y-2}^{-\!1}\! \cdot\! ...\! \cdot\! \sigma_{i+\!1}^{-\!1}\! \cdot\! \sigma_{i}^{-\!1})^{-\!1},\\
\end{array}
\end{equation}
\end{widetext}

\noindent with 1DURs (for $\alpha =\pi$) $ e^{iq\pi}$ (for $+$) and $e^{i(q-2)\pi}$ (for $-)$ (with supplement of the above notation for $x(y)=1$, 
$\sigma_i\cdot\sigma_{i+1}\cdot\dots \cdot\sigma_{i+x-2}\cdot \sigma_{i+x-1}
\cdot \sigma_{i+x-2}^{-1}\cdot \dots \cdot\sigma_{i+1}^{-1}\cdot\sigma_{i}^{-1}=\sigma_i$).
Examples of these braid generators are depicted in Fig. \ref{fig3}. 

The related modification of the Jastrow polynomial in the Laughlin function (\ref{666}) must be thus as follows (in the LLL the true ground state  wave function must be holomorphic function uniquely defined by its nodes):

\begin{widetext}
\begin{equation}
\label{888}
\begin{array}{l}
\Psi^{x,y,+}_q(z_1,z_2,\dots, z_N)=A\prod\limits_{\scriptscriptstyle{i,j=1;i<i\;\operatorname{mod}\;x + (j-1)x}}^{N,N/x}(z_i-z_{\scriptscriptstyle{i\;\operatorname{mod}\;x + (j-1)x}})^{q-1}\prod\limits_{\scriptscriptstyle{i,j=1;i<i\;\operatorname{mod}\;y + (j-1)y}}^{N,N/y}(z_i-z_{\scriptscriptstyle{i\;\operatorname{mod}\;y + (j-1)y}}) e^{-\sum\limits_i^N\frac{|z_i|^2 }{4l_B^2}},\\
\Psi^{x,y,-}_q(z_1,z_2,\dots, z_N)=A\prod\limits_{\scriptscriptstyle{i,j=1;i<i\;\operatorname{mod}\;x + (j-1)x}}^{N,N/x}(z_i-z_{\scriptscriptstyle{i\;\operatorname{mod}\;x + (j-1)x}})^{q-1}\prod\limits_{\scriptscriptstyle{i,j=1;i<i\;\operatorname{mod}\;y + (j-1)y}}^{N,N/y}(z_{\scriptscriptstyle{i\;\operatorname{mod}\;y + (j-1)y}} - z_i) e^{-\sum\limits_i^N\frac{|z_i|^2 }{4l_B^2}}.\\
\end{array}
\end{equation} 
\end{widetext}

\noindent The above functions for conventional composite fermion  hierarchy ($x=1$) attain the form (on the other hand, they define in a unique manner the unclear projection onto LLL in the conventional model \cite{jain2007}),  

\begin{widetext}
\begin{equation}
\label{999}
\begin{array}{l}
\Psi^{x=1,y,+}_q(z_1,z_2,\dots, z_N)=A\prod\limits_{\scriptscriptstyle{i,j=1,i<j}}^{N,N}(z_i-z_j)^{q-1}\prod\limits_{\scriptscriptstyle{i,j=1;i<i\;\operatorname{mod}\;y + (j-1)y}}^{N,N/y}(z_i-z_{\scriptscriptstyle{i\;\operatorname{mod}\;y + (j-1)y}}) e^{-\sum\limits_i^N\frac{|z_i|^2 }{4l_B^2}},\\
\Psi^{x=1,y,-}_q(z_1,z_2,\dots, z_N)=A\prod\limits_{\scriptscriptstyle{i,j=1,i<j}}^{N,N}(z_i-z_j)^{q-1}\prod\limits_{\scriptscriptstyle{i,j=1;i<i\;\operatorname{mod}\;y + (j-1)y}}^{N,N/y}(z_{\scriptscriptstyle{i\;\operatorname{mod}\;y + (j-1)y}} - z_i) e^{-\sum\limits_i^N\frac{|z_i|^2 }{4l_B^2}}.\\
\end{array}
\end{equation} 
\end{widetext}

\noindent The functions (\ref{888}) are proposed as the trial wave functions for correlated states for filling rates (\ref{444}) for which elementary exchanges of particles are defined by braids (\ref{777}) and generalize the Laughlin function (\ref{666}) for the case when $x,y>1$ with  some resemblance  to multicomponent Halperin functions \cite{halp}. 

The energy gain in the Laughlin state is due to lowering of the Coulomb repulsion energy $\left<\Psi\right|\sum\limits_{i.j,i>j}^{N,N}\frac{e^2}{|z_i-z_j|}\left|\Psi\right>$. It is clear that the energy reduction with the function 
(\ref{888}) is the weaker the higher $x$ is (for the same $q$ and $y$). It follows from the dilution of correlated particles for $x>1$ (the correlation concerns every $x$th electron only) as expressed in modified Laughlin-type function (\ref{888}) by reduction of the domain of the product. This leads to the diminishing of the repulsion energy gain due to the averaging of the Coulomb energy, $\sum\limits_{i.j,i>j}^{N,N}\frac{e^2}{|z_i-z_j|}$, with the wave function (\ref{888}) instead of (\ref{999}) (or (\ref{666})) because $q-1$ fold zero in these functions prevents approaching not all electrons in the case of function (\ref{888}) but only its $1/x$ fraction (opposite to the case of function (\ref{666}) or (\ref{999}) for which $x=1$). Therefore states with lower $x$ are more stable. Thus states with $x=1$ energetically prevail over states with $ x>1$ and are more stable. To confront the energy values obtained from exact diagonalization for different FQHE fillings \cite{jain-exact-2015}, the numerical estimation of energy  for newly proposed functions (\ref{888},\ref{999}) was  performed according to the Monte Carlo  Metropolis scheme \cite{montecarlo1,montecarlo2,metropolis}. Some exemplary results revealing very good overlap with the exact diagonalization  are presented in Table \ref{tab2}.

\begin{table}[th]
\centering
\begin{footnotesize}
\begin{tabular}{p{0.2cm}|p{0.2cm}|p{0.2cm}|p{2.1cm}|p{3cm}|p{2.2cm}}
$q$ & $x$ & $y$ & hierarchy fraction, $\nu = N/N_0$ & energy from Monte Carlo simulation for functions according to Eq. (\ref{888},\ref{999}) & energy from exact diagonalization \cite{jain-exact-2015}\\
\hline
3&1&2&$\frac{2\cdot 1}{(3-1)\cdot 2+1}=\frac{2}{5}$&$-0.432677$&$-0,432804$\\
\hline
3&1&3&$\frac{3\cdot 1}{(3-1)\cdot 3+1}=\frac{3}{7}$&$-0.441974$&$-0,442281$\\
\hline
3&1&4&$\frac{4\cdot 1}{(3-1)\cdot 4+1}=\frac{4}{9}$&$-0.446474$&$-0,447442$\\
\hline
3&1&5&$\frac{5\cdot 1}{(3-1)\cdot 5+1}=\frac{5}{11}$&$-0.451056$&$-0,450797$\\
\hline
5&1&2&$\frac{2\cdot 1}{(5-1)\cdot 2+1}=\frac{2}{9}$&$-0.342379$&$-0,342742$\\
\hline
5&1&3&$\frac{3\cdot 1}{(5-1)\cdot 3+1}=\frac{3}{13}$&$-0.348134$&$-0,348349$\\
\hline
5&1&4&$\frac{4\cdot 1}{(5-1)\cdot 4+1}=\frac{4}{17}$&$-0.351857$&$-0,351189$\\
\end{tabular}
\end{footnotesize}
\caption{Comparison of energy values obtained by exact diagonalization and by Monte Carlo simulation for some exemplary filling fractions for FQHE (Monte Carlo Metropolis simulation for the proposed topology-based wave functions, for 200 particles).} 
\label{tab2} 
\end{table}

Nevertheless, it should be commented that from the point of view of the commensurability condition governing the form of the cyclotron braid generator corresponding to multiloop cyclotron orbits, none of the loop cannot be featured, thus each loop can be accommodated to the particle separation independently. Thus, for $q$-looped orbit one would deal with the ordered series $x_1\leq x_2\leq \dots \leq x_q$ simplified in (\ref{444}) to $x_1=\dots=x_{q-1}=x, \; x_q=y$. Apparently, the Coulomb repulsion minimization prefers $x_1=\dots=x_{q-1}$ for which the minimization domain restriction (resulting in weaker interaction energy reduction) is more convenient than for distinct distributions of $x_i$. This explains the choice of the uniform behavior od $q-1$ loops (i.e., $x_1=\dots =x_{q-1}=x$) but this is not a rule and for many fractions various energetically competitive commensurability opportunities might be considered. 

The another observation related to various types of correlation identified by the commensurability criterion agrees with experimental data for the longitudinal resistivity $R_{xx}$ \cite{pan2003}, which is zero for states with all correlated particles (i.e., with $x=1$), whereas the residual its value grows with $x>1$ probably due to scattering on portion of non-correlated electrons. 

\section{Summation over configurations in the Feynman path-integral for a nonstationary problem}

The presented above identification of the ground state upon the path-integral scheme in Hall system at particular filling rate is the stationary problem with the eigen-energy well defined. This energy  may be numerically estimated, e.g., by the Metropolis Monte Carlo method (which has been illustrated above) as an expectation value for the suitable to the braid symmetry chosen trial wave-function. If the trial wave-function actually is the ground state function then the energy expectation value is the ground state energy. It has been proved that the Laughlin function is the true ground wave function when the Coulomb interaction is confined to near-range part, when so-called Haldane psudopotential terms \cite{haldane}, being matrix elements of the Coulomb interaction in terms of the relative angular momentum $m$ of the electron pair, are neglected for $m\geq q$, $q$ is an exponent in the Jastrow polynomial in the Laughlin state. Its closeness to a true ground state is confirmed by exact diagonalization in small models with accuracy 99\%. Similar accuracy confirms other states as indicated in Tab. \ref{tab2}.  

We have, however, noticed that for the same filling factor there are several commensurability instances and thus there are several related candidates for the ground state  trial wave-function at this filling rate. These functions differ in expectation energy and may be considered as excited states  with respect to that one  for which the energy is minimal. The excitation would be here associated with a reorganization of the correlation. The choice of the minimal energy and the corresponding wave function  is the picture of a stationary problem corresponding to  experimental measurements of conserved quantum quantities, as, in particular, an activation energy. In experiments an exact energy as the quantum number  is not usually accessible  but rather the thermodynamic its average and only due to extremely low temperature in Hall experiments one can refine the particular averaged energy value exceeding chaos contributing to the Gibbs distribution.

\begin{figure*}[ht]
\centering
\resizebox{0.8\textwidth}{!}{\includegraphics{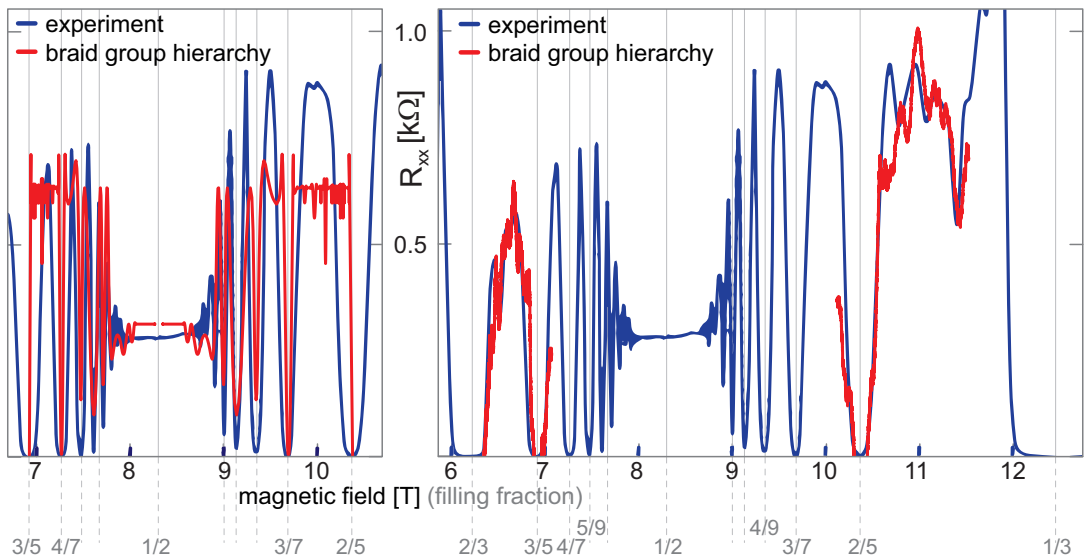}}
\caption{\label{rys1000} Close similarity of the function displaying the relative number of different topological configurations (red line) with the experimental curve for longitudinal resistivity $R_{xx}$ in the LLL for GaAs 2DEG (blue line---after Ref. \onlinecite{pan2003}).}
\end{figure*}

However, when we deal with nonstationary effects, like transport phenomena, the energy is not a conserved quantum number and its expectation value is averaged both quantumly over a non-stationary state and thermodynamically over the Gibbs distribution. If the energy is not determined then all excited stationary states corresponding to different commensurabilities (at the same fixed filling fraction must contribute to a non-stationary effect. This property influences significantly the Feynman path-integral which describes  as well non-stationary effects as stationary ones depending on whether the evolution operator has the integrand  with  the action explicitly time-dependent or not \cite{feynman1964}. A non-stationary problem  fits to e.g., a transport of charge in  an applied lateral electric field. In particular the longitudinal  conductivity (equal in 2D  to the  resistivity and is conventionally measured in Hall experiments \cite{prange}) is such a non-stationary quantity and must be proportional to suitable  path-integral in the configuration space between contact points separated by $l$ and time interval $l\rho /|\textbf{j}|$ ($l$ length of the sample, $\textbf{j}$ current density, $\rho$ electron density). One can argue that the value of the path-integral will be thus additionally  summed over all possible  different topological configurations at fixed filling ratio.  
For each configuration we deal with a distinct commensurability instance for multi loop braids and, hence, with a distinct cyclotron subgroup for the summation over homotopy classes in the final path-integral. In the case of a nonstationary path-integral all configurations will interfere and must be taken into account by summation. The latter does not cause any conflict between statistics as for all cyclotron subgroups for original fermions the 1DURs are given by $e^{p\pi}=-1$ where $p$ is an odd integer (as shown in  Sec. \ref{3}). Hence, the summation over configurations approximately resolves to the multiplication of the path-integral by the number of various configurations (neglecting a normalization). As the various filling factors admit a different number of distinct commensurability instances, hence one can compare an increase of the path-integral at varying filling rate. The longitudinal conductivity measured in Hall systems is proportional to the propagator in position representation, hence variation of the path-integral with changing filling factors ought to be visible in the longitudinal conductivity. In Fig. \ref{rys1000} we plot the relative number of various configurations with respect to the filling factors (the red line). We notice its similarity with the experimentally measured conductivity curve---marked as the blue line in Fig. \ref{rys1000}. Note, that the demonstrated behavior of $R_{xx}$, e.g., in the $\nu=\frac{1}{2}$ vicinity,  has not been explained previously.

\section{Bohr-Sommerfeld quantization rule in homotopy-rich 2D space}
\label{10}

Let us consider a quasiclassical function $\Psi=Ce^{iS/\hbar}$. If one takes into account  two first terms of the Schr\"odinger equation with respect to powers of $\hbar$, then one arrives with the quasiclassical formula for the stationary state in an  arbitrary 1D well, $U(x)$ with turning points $a$ and $b$, $\Psi(x)=\frac{c}{\sqrt{p}} sin \frac{1}{\hbar}\int_a^xpdx$, for $\Psi(a)=0$  or $\Psi(x)=\frac{c'}{\sqrt{p}} sin \frac{1}{\hbar}\int_b^xpdx$, for $\Psi(b)=0$, where $p(x)=\sqrt{2m(E-U(x))}$. From the wave-function uniqueness requirement, $2 \int_a^bpdx=  \oint p dx=  S_{xp}= n 2\pi \hbar =n h$. This is the Bohr-Sommerfeld quantization rule. 

However, if $(ab)$ trajectory in the classical phase space may have some odd topology, then 2 $ \int_a^bpdx= \oint pdx= S_{px}= (2k+1) n 2\pi \hbar =n (2k+1) h$, for a trajectory  $(a,b)$ with additional $k$ loops. Such a trajectory may be in 2D  non homotopic, in general, with the trajectory $(a,b)$ without any additional loops, when was $ 2\int_a^bp/\hbar dx= \oint p/\hbar dx =  n 2\pi$.  Each loop of all $2k$ loops symmetrically pinned  (by $k$) to  both branches, 'upper' ($+p$) and 'lower' ($-p$), of the closed trajectory between $a$ and $b$  adds $2\pi$, which gives $2k+1$ final factor. In general,  one must thus take into account the possible homotopy  oddness of the 2D phase space topology.
In the case of the ordinary  phase space (2D) of 1D particle such an oddness does not happen, but when the Bohr-Sommerfeld rule is applied to a fictitious 2D phase space ($P_y,Y$) of $y,x$ components of the 2D kinematic momentum in the presence of a magnetic field, $P_x=-i\hbar\frac{\partial}{\partial x}$ and $P_y=-i\hbar \frac{\partial}{\partial y} -eBx  $  (at the Landau gauge, $\textbf{A}=(0,Bx,0)$),  $[P_y,P_x]_-=-i\hbar eB$, then the 2D fictitious phase space $(Y,P_y)$ ($Y=\frac{1}{eB}P_x$, $[P_y,Y]_-=-i\hbar$) is actually the $(P_x,P_y)$ space, which is renormalized by factor $\frac{1}{(eB)^2}$ and turned in plane by $\pi/2$ the  ordinary 2D  space $(x,y)$ (as $dP_{x(y)}=eB dy(-x)$ due to Lorentz force). In this $(x,y)$ space trajectories may not be homotopic and may be assigned with non-contractible additional loops (as in  multiparticle planar systems). 

Hence, in this homotopy-rich 2D case  we obtain  the generalized Bohr-Sommerfeld rule, $S_{YP_y}= n (2k+1) h$, or in $(x,y)$ space, $\Delta S_{xy} B=\frac{(2k+1)h}{e}$.

IQHE corresponds to $k=0$, $\Delta S_{xy}= \frac{h}{eB}= \frac{S}{N}=\frac{S}{N_0}$, $\nu=\frac{N}{N_0}=1$ ($N_0=\frac{BS}{h/e}$ is the LL degeneracy, $S$ is the sample surface size, $N$ is the number of electrons).

FQHE corresponds to $k=1,2,\dots$, and e.g., for $k=1$ (Laughlin state), $\Delta S_{xy}= \frac{3h}{eB}=\frac{S}{N}$ and hence $\nu=\frac{N}{BSe/h}=\frac{1}{3}$.

The quasiclassical method of Bohr-Sommerfeld in 2D Hall systems is interaction independent  (it holds  for both non-interacting and interacting systems), though an existence of non homotopic trajectories in $(x,y)$ space is conditioned by the Coulomb interaction of 2D charged particles. This approach confirms our previous estimation of the size of the multi loop cyclotron orbits and proves that the cyclotron orbit size grows with number of loops $k$, $\Delta S_{xy}B=\frac{(2k+1)h}{e}$. The proved above property means that  in 2D for multi loop ($k$-loop) trajectories the magnetic flux quantum is $\frac{(2k+1)h}{e}$ (and  $\frac{h}{e}$ only for $k=0$).

\section{Comments and conclusion}
We demonstrated that Feynman path-integrals are especially suitable for the analysis  of  multiparticle systems of indistinguishable   charged particles located on 2D manifold and subjected to a strong perpendicular magnetic field, when the path space is exceptionally homotopy-rich. The path-integral approach allows to directly incorporate the topology (homotopy) effects in distinction to local quantum mechanics upon the Schr\"odinger equation formulation. Unlike local quantum mechanics path integration method of quantization evokes classical trajectories in view of  the least action principle which is especially dedicated to include homotopy effects which are also classical ones and concern classical trajectories. Taking advantage of the famous Feynman formulation, the summation of the complex amplitudes expressed by a semiclassical wave function for each trajectory joining distant points in the configuration space at some time interval results in an interference of this complex amplitudes. As a rule in quantum mechanics there are summed up  complex amplitudes of probabilities but not probabilities themselves and the sum of  amplitudes, $e^{iS/\hbar}$, gives the evolution operator matrix element in position representation between assumed initial and final positions for a selected time interval.  Trajectories can be, however, classified in terms of the homotopy $\pi_1(A)$ group of the configuration space $A$. When the system consists of many identical indistinguishable particles then its configuration space is multiply connected and $\pi_1(A)$ is nontrivial in contrary to a simply connected configuration space of a single-particle with $\pi_1(A)$ trivial. The $\pi_1$ group for the  configuration space of $N$ indistinguishable particles is called as the full braid group. Braids are loops in the configuration space and are topologically inequivalent. These loops may be adjoint to trajectories in the path-integral, which also must be inequivalent if various braids are adjoint. This breaks the condition of continuity for the measure definition for path integration, and contributions of non homotopic sectors of the path space must be integrate separately and summed over the full braid group with some unitary weight factors (to conserve causality). These weight factors form one dimensional unitary representation of the full braid group (1DUR), and  there exist as many quantum counterparts of initial classical particles as many different 1DURs exist. For 2D manifold were initial $N$ classical particles are located one can arrive with bosons, fermions or anyons.

We demonstrated, however, that in the presence of a strong magnetic field for 2D electrons the full braid group may considerably change. The change depends on the planar concentration of  uniformly distributed (in the form of Wigner crystal) Coulomb repulsing electrons. When particle separation fits to a size of the  cyclotron orbits of electrons then the braids exist, otherwise not. If single-loop cyclotron orbits are  shorter than electron separation, then it may happen that only multi loop orbits fit to the separation between electrons. multi loop cyclotron orbits in 2D have larger size than in 3D or 2D single-loop orbit, because exclusively  in 2D the flux passing the multi loop orbit must be divided between single loops. We have proved this effect by the application of the Bohr-Sommerfeld rule.
 
Various patterns of the commensurability between multi loop braids and electron separation including nearest and next-nearest electron neighbors result in various so called cyclotron braid subgroups of the full braid group. These cyclotron braid subgroups define thus  different  domains for summation over homotopy classes in the path-integral for varying magnetic field value (or equivalently, for varying filling rate of the Landau level). 1DURs for these subgroups define new quantum particles (beyond bosons, fermions, anyons) which we call composite fermions (the name, by historical reason to link with a conventional Jain's composite fermion model), composite bosons or composite anyons. These homotopy induced composite fermions explain in all details the experimentally observed hierarchy in FQHE, which was inaccessible for the local quantum mechanics and the conventional composite fermion model. 

The Feynman path-integral method came out to be also very helpful in another problem of homotopy-rich multiparticle systems. We have demonstrated that in the case of a non-stationary problem, when the energy is not a quantum number, the path-integral can be also utilized with the action $S$ explicitly time dependent in contrary to the stationary case. In the non-stationary case the space of paths is more complicated in comparison to the stationary case. In the homotopy-rich system, like 2D electrons upon strong magnetic field at a fixed filling rate, various patterns of commensurate multi loop braids are possible with different energies and additional summation over all these patterns must be included in the full domain of the path-integral for any non-stationary problem. We have proved that an inclusion of such a summation over configurations (i.e., over various commensurability patterns at the same filling rate) results in a very good consistence of the relative values of the path-integral and relative values of the longitudinal conductivity measured in FQHE versus the filling rate, because the number of configurations strongly depends on the filling rate. 
   
The above described two effects due to the inclusion of homotopy classes into path integration evidence the high significance of the Feynman path integration approach to multiparticle quantum systems. This approach not only enhances transparency of the theory and understanding of the related quantum  behavior   but also displays effects not noticed with the conventional local quantum mechanics.

\begin{acknowledgments}
Supported by the NCN projects P.2011/02/A/ST3/00116 and P.2016/21/D/ST3/00958.
\end{acknowledgments}

\end{document}